\documentstyle[12pt,psfig]{article}

\begin{document}
\pagestyle{empty}
\begin{flushright}
{BROWN-HET-962} \\
{DAMTP 94-72} \\
{PUPT-94-1497} \\
{September 1994}
\end{flushright}
\vspace*{5mm}
\begin{center}
{\bf LOCAL AND NONLOCAL DEFECT-MEDIATED ELECTROWEAK BARYOGENESIS} \\
[10mm]
\renewcommand{\thefootnote}{\alph{footnote}}
Robert Brandenberger$^{1,2,}$\footnote{e-mail: rhb@het.brown.edu.},
Anne-Christine Davis$^{1,3,}$\footnote{e-mail:
A.C.Davis@damtp.cambridge.ac.uk.},
Tomislav Prokopec$^{1,4,}$\footnote{e-mail: prokopec@puhep1.princeton.edu},\\
Mark Trodden$^{1,2,}$\footnote{e-mail: mtrodden@het.brown.edu.}\\
[10mm]

\end{center}
\begin{flushleft}
1){\it Isaac Newton Institute for Mathematical Sciences \\
University of Cambridge, Cambridge, CB3 OEH, U.K.}\\
2){\it Physics Department, Brown University,Providence RI. 02912. USA.} \\
3){\it Department of Applied Mathematics and Theoretical Physics and Kings
College, University of Cambridge, Cambridge CB3 9EW. U.K.}\\
4){\it Joseph Henry Laboratories, Princeton University, Princeton, NJ. 08544.
USA.}
\end{flushleft}
\begin{center}
{\bf Abstract}
\end{center}
\vspace*{3mm}
We consider the effects of particle transport in the topological
defect-mediated electroweak baryogenesis scenarios of Ref. 1. We analyze
the cases of both thin and thick defects and demonstrate an enhancement of
the original mechanism in both cases due to an increased effective volume in
which baryogenesis occurs. This phenomenon is a result of imperfect
cancellation between the baryons and antibaryons produced on opposite
faces of the defect.
\setlength{\textheight}{8.5in}
\newpage\setcounter{page}{1}\pagestyle{plain}
\renewcommand{\thefootnote}{\arabic{footnote}}
\section{Introduction}

Since the realization \cite{KRS 85} that the standard model of electroweak
interactions contains all the ingredients neccessary to explain the baryon
asymmetry of the universe (BAU) there have been many attempts to show in
detail how such an asymmetry may be dynamically generated in this context
(see Refs. 3
and 4 for recent reviews).

The neccessary conditions to generate a net baryon number are\cite{Sakharov}:
\begin{enumerate}
\item the existence of baryon number violating processes,
\item C and CP violation,
\item departure from thermal equilibrium.
\end{enumerate}

Almost all presently proposed electroweak baryogenesis scenarios achieve the
first requirement by using finite temperature sphaleron transitions
\cite{{NM 83},{GtH 76}}, the second by using an extended, CP-violating
Higgs sector for the standard model (C is violated maximally in the
Weinberg-Salam theory) and the third by requiring that the electroweak
phase transition be sufficiently strongly first order. If this final
assumption holds then the phase transition proceeds via bubble nucleation
and baryogenesis takes place in the bubble walls where the Higgs fields
cause the departure from thermal equilibrium.

In a recent paper \cite{BDT 94} three of us (RB, ACD, MT)  suggested an
alternative realization of the third Sakharov condition in the context of
the electroweak phase transition in the presence of topological defects
remaining after a previous symmetry breaking. The electroweak symmetry is
restored out
to some distance around these defects and, as segments of them collapse,
the loss of thermal equilibrium caused by the transition from false to
true
vacuum allows local baryogenesis to occur at the outer edge of the defect.
An estimate of the baryon to entropy ratio produced by this mechanism was
performed and compared to the strength of previous mechanisms. Its strength
was found to be suppressed by  the ratio of the volume in defects to the
total volume. For defects formed at scales close to the electroweak scale
and for optimistic values of the parameters the mechanism was shown to
give results consistent with the observed baryon to entropy ratio.

The advantage of such a scenario for electroweak baryogenesis is that it
does not depend in any way on the order of the electroweak phase transition.
This is a significant advantage, given that the order of the phase transition
is not known at present, and that in particular for large Higgs mass the
transition is unlikely to proceed via the nucleation of critical bubbles. A
further advantage of using topological defects to seed baryogenesis is that
the volume in defects decreases only as a power of time below the phase
transition temperature. Therefore, as pointed out in Ref. \cite{BDH 91},
defect-mediated baryogenesis is still effective even if sphalerons are in
thermal equilibrium just below the electroweak phase transition temperature.
However, a potential drawback for specific implementations is the
requirement that defects be formed at a scale rather close to the
electroweak scale in order to avoid large volume suppression factors.

In this paper we attempt to relax this requirement by performing a more
detailed analysis of the mechanism. In particular we show that the volume
suppression discussed above may be relaxed when the imperfect cancellation
between the baryons and antibaryons produced is considered. We also
demonstrate that not only local baryogenesis \cite{{TZ 90},{CKN1 91}}
(where sphaleron transitions and CP violation take place at the same
spacetime point) but also nonlocal baryogenesis \cite{{CKN2 92},{JPT2 94}}
(where the sphaleron transitions and CP violation act in different regions
of space: CP violation in the bubble or defect wall leads to asymmetries
in quantum numbers other than baryon number which are then converted to
baryon number by sphaleron processes in the larger region where the symmetry
is restored) is important. These particle transport processes lead to an
increase in the sphaleron transition rate (since the latter is unsuppressed
in the false vacuum) and to a considerable enhancement of the local baryon
number density generated and thus to an improved estimate of the net baryon
asymmetry capable of being produced by this mechanism.

The outline of this paper is as follows. In section two we review our previous
work and outline the details of our new approach. Section three is devoted to
an analysis of local baryogenesis as considered in Ref.~1 but with the new
volume enhancement factors taken into account. In section four we describe
how nonlocal baryogenesis can enhance the number density of baryons produced
by defects. This enhancement can take either of two forms depending on the
thickness of the walls of the defects. Section five contains a discussion of
the effects of different geometries (cosmic strings, domain walls, closed
and infinite defects) and types of motion (collapse and translational) on
the resulting BAU. In particular, we examine several detailed examples of
the scenario, one of which may be directly compared with our previous
estimate. Finally, in section six, we conclude and discuss our results.

\section{Enhancement of the Baryogenesis Volume}
Let us briefly review the scenario for electroweak baryogenesis proposed in
Ref~1.

Consider a $d$-dimensional topological defect produced at an energy scale
$\eta > \eta_{\rm EW}$, where $\eta_{\rm EW}$ is the electroweak scale. In
this work we shall only be interested in one dimensional defects - cosmic
strings - and two dimensional defects - domain walls.

Assume that there is some process whereby baryon number is violated (e.g.
sphaleron-induced transitions). Further, assume that the rate $\Gamma$ per
unit volume for this process is given by
\begin{equation}
\Gamma = \left\{ \begin{array}{ll}
         \Gamma_B(T) & \ \ \ \mbox{$T>\eta_u$} \\
         0           & \ \ \ \mbox{$T<\eta_u$}
         \end{array} \right.
\end{equation}
where $\eta_u \leq \eta_{\rm EW}$ is the energy scale at which sphaleron
transitions in the plasma external to the string become exponentially
suppressed. If certain consistency conditions are satisfied (eg. sphalerons
fit inside the defect so that their rate is not significantly suppressed)
then within the defect the rate of baryon number violation is
$\Gamma_B(T_{\rm EW})$ after the electroweak phase transition.

Also, assume that in the Higgs sector of the electroweak theory there
exist CP-odd terms (for a recent attempt to enhance CP violation in the
standard model by means of a condensate of a CP-odd Z field on the bubble
wall see \cite{NT 94}). In this case, extra CP violation not present in
the standard model can take place inside the defect walls. We are interested
in the relative production of baryons and antibaryons due to the motion of
these topological defects (see \cite{{BDH 91},{BD 93}} for other ways to
use topological defects to generate the baryon to entropy ratio).

To be definite we shall assume that the CP violation is due to a CP-odd
relative phase, $\theta$, between the two electroweak Higgs doublets - this
may be seen as the restriction that both Higgs may not simultaneously take
real vacuum expectation values - and that this phase changes by
$\Delta\theta_{CP}$ during the transition from false to true vacuum, and by
$ - \Delta\theta_{CP}$ in the reverse transition.

In the electroweak theory baryon number is an anomalous global symmetry
\cite{{GtH 76},{Arnold Review}}. This is the origin of the baryon number
violation. There are several mechanisms by which the CP-odd phase
$\theta_{\rm CP}$ can contribute to the free energy density of the theory.
To be specific, we shall concentrate on a one-loop effect (see for
example \cite{{TZ 90},{ACW 93}}). Tree level effects have been discussed
in Refs. \cite{CKN1 91} and \cite{JPT3 94}. The one loop contribution is
\begin{equation}
{\cal F}_B = -\frac{14}{3\pi^2 N_f}\zeta(3)\left(\frac{m}{T}\right)^2{\dot
\theta}_{\rm CP} n_B
\end{equation}
where $m$ is the (finite temperature) mass of the particle species dominating
the contribution to the anomaly and $\zeta$ is the Riemann $\zeta$-function.
The coefficient of $n_B$ in the above equation can be viewed as a chemical
potential $\mu$
\begin{equation}
\mu = \frac{14}{3\pi^2 N_f}\zeta(3)\left(\frac{m}{T}\right)^2{\dot \theta}_{\rm
CP}
\end{equation}
for baryon number. Thus, it is clear that if $\Delta \theta_{\rm CP} >0$ for a
given process then baryon number is driven positive (an excess of baryons over
antibaryons is generated) and vice-versa.

As a defect moves, certain regions of the background space enter the core of
the defect - i.e. make the transition from true to false vacuum - while others
leave the core and make the transition from false to true vacuum. In Ref. 1
it was shown that certain types of motion and evolution of defects can provide
an asymmetry such that an overall baryon excess is created in the universe.

In the original work of Ref.~1 the processes responsible for the generation of
the baryon asymmetry were purely local. By this we mean that the baryon number
violating interactions and the CP violation neccessarily take place at the same
spacetime point. This requirement leads to restrictions on the strength of the
mechanism since an important quantity which enters the calculation is the
suppression factor
\begin{equation}
({\rm SF}) \sim \left(\frac{V_{\rm BG}}{V}\right)
\end{equation}
where $V_{\rm BG}$ is the volume in which baryogenesis occurs and $V$ is the
total volume. $(SF)$ is the factor by which defect-mediated baryogenesis is
weaker than baryogenesis with bubble walls. For a collapsing topological
defect, purely local baryogenesis restricts the baryogenesis volume to be the
initial volume of the defect because the effects of ${\dot \theta}>0$ on one
side of the string are cancelled by the effects of ${\dot \theta}<0$ on the
other. In order for $(SF)$ not to be a prohibitively small suppression it was
neccessary for us to require that the scale at which our defects are formed be
close to the electroweak scale.

Our aim here is to investigate the mechanism in much more detail. In particular
we wish to take into account the various types of particle interactions within
the string, where the symmetry is unbroken. These interactions allow us to
enhance the local effects in Ref.~1. because now the decay of antiparticles
produced by the leading edge of the defect results in imperfect cancellation
between the effects of the competing processes.

We shall also examine the effects of nonlocal baryogenesis which allow us to
further increase the number of baryons which our mechanism produces. This may
take one of two forms:
\begin{enumerate}
\item
Particle reflection \cite{{CKN2 92},{JPT2 94}}: particles within the defect
reflect preferentially off the advancing face of the defect due to CP violation
and result in a net chiral flux which propagates back into the interior of the
defect and is converted to baryon number. This process is efficient only for
``thin" walled defects.
\item
Classical force \cite{{JPT3 94},{CKN 94}}: as a result of CP violation, an
axial field emerges on the wall leading to a classical force which perturbs
particle densities, thus biasing baryon number. When the effects of particle
transport are taken into account, this leads to nonlocal baryogenesis. This
process dominates for rather thick walls.
\end{enumerate}
Both these mechanisms lead to an increase in the net effective volume
contributing to baryogenesis over that of local baryogenesis since we no longer
rely on anomalous interactions taking place in the narrow region of the face of
the defect where the changing Higgs fields provide CP violation.

The chiral asymmetry which is converted to an asymmetry in baryon number is
carried by both quarks and leptons. However, since the Yukawa couplings of the
top quark and the $\tau$-lepton are larger than those of the other quarks and
leptons respectively, we expect that the main contribution to the reflected
asymmetry will come from these particles and henceforth we ignore the effects
of the other particles.

When considering nonlocal baryogenesis it is convenient to write the equation
for the rate of production of baryons in the form \cite{JPT2 94}
\begin{equation}
{\dot B} = -\frac{N_f\Gamma_s}{2T}\sum_i\mu_i
\end{equation}
where the rate per unit volume for electroweak spaleron transitions is
\begin{equation}
\Gamma_s=\kappa(\alpha_{\rm W}T)^4
\end{equation}
with \cite{{ALS 89},{GST 92}} $0.1 \leq \kappa \leq 1$, $N_f$ the number of
families and $\mu_i$ is the chemical potential for left handed particles of
species $i$. $B$ is the number density of baryons produced locally by the
process. The crucial question in applying this equation is an accurate
evaluation of the chemical potentials that bias baryon number production.

\section{Local Baryogenesis}

In this section we shall obtain a revised estimate for the baryon asymmetry
produced by a topological defect as a consequence of local mechanisms.

As a topological defect passes each point in space a number density of
antibaryons is produced by local baryogenesis at the leading face of the defect
and then an equal number density of baryons is produced as the trailing edge
passes. Naively we would expect that these effects would cancel each other, so
that any time-symmetric motion of the defect such as translation would yield no
net baryon asymmetry. This is the reason that in Ref.~1 we restricted ourselves
to a time-asymmetric motion -- loop collapse -- to generate a baryonic excess.
The cancellation effects led to the suppression of the strength of our
mechanism by the factor $(SF)$ mentioned in section~2.

However, in this treatment we have neglected an important effect and thus
underestimated the strength of the mechanism. The antibaryons produced at the
leading edge of the defect at a fixed point in space spend a time interval
$\tau$ inside the defect during which they may decay before the trailing edge
passes by and produces baryons at the same point. The core passage time $\tau$
is given by
\begin{equation}
\tau = \frac{L}{v_{\rm D}},
\end{equation}
where $L$ is the width of the defect and $v_{\rm D}$ is its velocity.

Thus, if $n_b^0$ is the number density of baryons (or antibaryons) produced at
either edge, we may estimate the net baryon asymmetry $B$ produced after the
defect has passed a given point once to be
\begin{equation}
B = n_b^0(1-e^{-{\bar{\Gamma}_s} \tau})
\end{equation}
where ${\bar{\Gamma}_s}$ is the rate at which antibaryons decay and may be
related to the electroweak sphaleron rate by \cite{{CKN2 92},{JPT2 94}}
\begin{equation}
{\bar{\Gamma}_s} = 6N_f \frac{\Gamma_s}{T^3} = 6N_f \kappa \alpha_{\rm W}^4 T.
\end{equation}

The resulting average baryon number density $n_b$ can be estimated from (3) and
taking into account (8) and the volume suppression (see (4)):
\begin{equation}
n_b \simeq 3 N_f {{\Gamma_s} \over {T}} \mu {{\delta} \over {v_D}} (1 - e^{-
{\bar \Gamma}_s \tau}) (SF),
\end{equation}
where $\delta$ is the thickness of the defect wall.
The derivative of $\theta_{CP}$ and $\delta / v_D$ combine to give $\Delta
\theta_{CP}$,
and hence the resulting net baryon to entropy ratio becomes
\begin{equation}
{{n_b} \over s} \simeq 4 \kappa \alpha_W^4 {g^*}^{-1} ({m \over T})^2 \Delta
\theta_{CP} {{V_{BG}} \over V}(1 - e^{- {\bar \Gamma}_s \tau}),
\end{equation}
where $g^*$ is the number of spin degrees of freedom which enters into the
equation for the entropy density $s = {{2 \pi^2} \over {45}} g^* T^3$.

In our original estimate the total volume contributing to baryogenesis was the
initial volume occupied by defects. Now, with this new effect taken into
consideration, this volume is dramatically increased. All the volume swept out
by the defect network participates in the effect.

We still have a volume suppression factor but its value is considerably larger
than before and, as we shall see in section~5, in some cases it is ${\cal
O}(v_{\rm D})$ because the defect network can sweep out that fraction of the
total volume in one Hubble expansion time.

Even if $V_{\rm BG}/V \sim 1$ we still have the suppression of this mechanism
over the usual bubble wall scenarios by the factor

\begin{displaymath}
(1-e^{-{\bar{\Gamma}_s} L/v_{\rm D}})
\end{displaymath}

This clearly distinguishes two cases. In the first case in which the defects
are ``thin" - defined as $L<v_{\rm D}/{\bar{\Gamma}_s}$ - we have a suppression
factor of approximately ${\bar{\Gamma}_s} L/v_{\rm D}$. If the defects are
``thick", $L>v_{\rm D}/{\bar{\Gamma}_s}$, then there is negligible suppression
due to this effect.

Let us now examine how these conditions are related to the microphysical
parameters of the models. First consider non-superconducting defects. The
electroweak symmetry is restored out to a distance \cite{PD 93}

\begin{equation}
R_s \sim \lambda^{-1/4} G^{-1/2}\eta_{\rm EW}^{-1}
\end{equation}
where $\eta_{\rm EW}$ is the electroweak scale, $G^2=g^2+g'^2$ and $g$ and $g'$
are the $SU(2)$ and $U(1)$ gauge couplings respectively. The defects are
considered ``thin" if the Higgs self coupling $\lambda$ satisfies

\begin{equation}
\lambda > \left(\frac{{\bar \Gamma}_s}{v_{\rm D}\eta_{\rm EW}}\right)^4
\frac{1}{G^2}
\end{equation}
and ``thick" otherwise. This quantity may be estimated by evaluating ${\bar
\Gamma}_s$ (see equation~9) at the electroweak temperature, $G\sim{\cal O}({1
\over {30}})$ and $v_{\rm D} \sim 0.1 -1$ resulting in the condition $\lambda
>10^{-23}-10^{-27}$, an inequality which includes most of the parameter space
of the theory. Thus, we may conclude that for the case of ordinary defects the
suppression factor ${\bar \Gamma}_s L/v_{\rm D}$ almost always applies.

Now consider the case where the defects are superconducting. If, as in Ref.~1,
we estimate the current on the defects by assuming a random walk of the winding
of the condensate field (assume scalar superconductivity for simplicity), then
we may estimate the size of the symmetry restoration region to be \cite{{PD
93},{ANO}}
\begin{equation}
R_s \sim \sqrt{\frac{1}{2\lambda}}\frac{1}{2\pi}\frac{1}{\eta_{\rm
EW}}\left(\frac{\eta}{\eta_{\rm EW}}\right)^{3/4}
\end{equation}
where $\eta$ is the scale at which the defects are formed. A similar result has
been shown to hold in the two-Higgs doublet model we are using \cite{MT 94}.
Thus, in this case, our defects are considered ``thick" if
\begin{equation}
\eta >\left(\frac{v_{\rm D} \eta_{\rm EW}}{{\bar \Gamma}_s
}\sqrt{2\lambda}2\pi\right)^{4/3} \eta_{\rm EW}
\end{equation}
and ``thin" otherwise. Using $\lambda \sim 1$ and estimating $\Gamma$ from (9)
we obtain $\eta > 10^8 - 4 \cdot 10^{10}$GeV.

Therefore, if the scale of the defects is in this range then there is no
additional suppresion beyond the volume suppression. If the scale lies below
this then we have the factor $\Gamma L/v_{\rm D}$ as in the case of ordinary
defects.

The above considerations allow us to compute the asymmetry in the baryon number
density at every point swept out by a topological defect of a given type. In
order to make a specific prediction we need to consider a particular type of
defect in a given configuration and have knowledge of the evolution of the
defect network. This then enables us to make a reliable estimate for the volume
suppression $(SF)$ and hence the total baryon asymmetry. In section 5 we shall
perform this calculation in some examples. Firstly, however, we shall examine
the mechanism when the baryon production is by nonlocal means.

\section{Nonlocal Baryogenesis}
We now turn to the issue of baryons produced by nonlocal mechanisms within the
defects. There are (at least) two distinct ways in which this can occur. One
mechanism applies when the walls of the defect, where the Higgs fields are
changing, are thin (in a sense which will be made precise in a moment) and the
other applies when the walls are thick.

Let us first consider the case where the Higgs fields change only in a narrow
region at the face of the topological defect. In analogy with the bubbles
formed during a first order phase transition we refer to this as the {\it thin
wall} case.

In this regime, effects due to local baryogenesis are heavily suppressed
because CP violating processes take place only in a very small volume in which
the rate for baryon violating processes is non-zero. However, we shall see that
nonlocal baryogenesis allows us to produce an appreciable baryon asymmetry due
to particle transport effects \cite{{CKN2 92},{JPT2 94}}.

In the rest frame of the topological defect particles within the core see a
sharp potential barrier and reflect off the trailing edge in a CP violating
manner due to the gradient in the CP-odd Higgs phase in the defect wall. The
same is true for particles reflecting back into the broken symmetry phase from
the leading edge of the defect. It is neccessary that the walls of the defect
be thin, defined as $\delta<l$ where $l$ is the mean free path of the relevant
particle species within the wall of thickness $\delta$, in order that coherent
quantum effects give unsuppressed reflection. A less restrictive condition
which may lead to an additional suppression factor is that the diffusion
constant, $D$, of the relevant species satisfy $D>\delta$ \cite{JPT2 94}.

After reflection the asymmetries in certain quantum numbers caused by the CP
violation diffuse in front of the respective face of the defect due to particle
interactions and decays \cite{{CKN2 92},{JPT2 94}}. In particular, the
asymmetric reflection of left and right handed particles will lead to a net
chiral flux from the wall. However, there is a qualitative difference between
the diffusion occuring in the two regions.

In the interior of the defect the electroweak symmetry is restored and weak
sphaleron transitions are unsuppressed. This means that the chiral asymmetry
carried into this region by the reflected particles may be converted to an
asymmetry in baryon number by sphaleron effects. In contrast, particles
reflected into the phase of broken symmetry may diffuse only by baryon number
conserving decays since the electroweak sphaleron rate is exponentially
suppressed in this region. Hence, we shall concentrate only on those particles
reflecting into the interior of the defect. In figure~1 we represent the
various processes on a diagram of the defect core.

\begin{figure}
\caption{Diagram of a portion of a defect, in this case a cosmic string, moving
to the right through the primordial plasma. The differing decays of reflected
particles within and outside the defect leads to the generation of a net baryon
asymmetry.}
\end{figure}

The net baryon to entropy ratio which results via nonlocal baryogenesis in the
case of thin walls can be calculated following the analyses in Refs. \cite{CKN2
92} and \cite{JPT2 94}. The baryon density produced by a single defect is given
by Eq. (5) in terms of the rate of baryon number violating processes - in turn
given by (6) - and the chemical potentials $\mu_i$ for left handed particles.
These chemical potentials are a consequence of the asymmetric reflection off
the walls and the resulting chiral particle asymmetry.

In the following we give a brief outline of the logic of the calculation. For
details see Refs. \cite{CKN2 92} and \cite{JPT2 94}. Baryon number violation is
driven by the chemical potentials for left handed leptons or quarks. We here
focus on leptons \cite{JPT2 94} (for
quarks see e.g. Ref. \cite{CKN2 92}). If there is local thermal equilibrium in
front of the defect walls - as we assume - then the chemical potentials $\mu_i$
of particle species $i$ are related to their number densities $n_i$ by
\begin{equation}
n_i = {{T^2} \over {12}} k_i \mu_i,
\end{equation}
where $k_i$ is a statistical factor which equals $1$ for fermions and $2$ for
bosons. It is important to correctly \cite{JPT1 94} impose the constraints on
quantities which are conserved in the region in front of and on the wall, but
at the level of this discussion we do not need to address this point.

Using the above considerations, the chemical potential $\mu_L$ for left handed
leptons can be related to the left handed lepton number densities $L_L$. These
are in turn determined by particle transport. The source term in the diffusion
equation is the flux $J_0$ resulting from the asymmetric reflection of left and
right handed leptons off the defect wall.

The asymmetric reflection coefficients for lepton scattering is
\begin{equation}
{\cal R}_{L \rightarrow R} - {\cal R}_{R \rightarrow L} \simeq 2 \Delta
\theta_{CP} {{m_l^2} \over {m_H \vert p_z \vert}}, \ \ \ \ \  m_l < \vert p_z
\vert < m_H \sim {1 \over \delta}
\end{equation}
where $m_l$ and $m_H$ are the lepton and Higgs masses, respectively, and $\vert
p_z \vert$ is the momentum of the lepton perpendicular to the wall (in the wall
frame). The resulting flux of left handed leptons is
\begin{equation}
J_0 \simeq {{v_D m_l^2 m_H \Delta \theta_{CP}} \over {4 \pi^2}},
\end{equation}
where $v_D$ is the defect translational velocity. Note that in order for the
momentum interval in Eq. 17 to be nonvanishing, the condition $m_l \delta < 1$
needs to be satisfied.

The general structure of the diffusion equation for a single particle species
is
\begin{equation}
D_L L_L^{\prime \prime} + v_D L_L^{\prime} = \xi^L J_0 \delta^{\prime}(z),
\end{equation}
where $D_L$ is the diffusion constant for leptons, $\xi^L$ is the persistence
length of the current in front of the defect wall, and a prime denotes the
spatial derivative in direction $z$
perpendicular to the wall. This equation can easily be solved by the Green's
function method, yielding
\begin{equation}
L_L(z) = J_0 {{\xi_L} \over {D_L}} e^{- \lambda_D z}
\end{equation}
where $\lambda_D = v_D / D_L$.

Finally, the chemical potential $\mu_L$ can be related to $L_L$ by
\begin{equation}
\mu_L = { 6 \over {T^2}} L_L
\end{equation}
(for details see Ref. \cite{JPT2 94}).

Inserting into Eq. (5) the sphaleron rate (6) and the above results for the
chemical potential $\mu$, and taking into account the suppression factors (4)
and the analog of (8) for nonlocal baryogenesis, we obtain the final baryon to
entropy ratio
\begin{equation}
{{n_b} \over s} = {{n_b^{(0)}} \over s} (1 - e^{- L \lambda_D}) {{V_{BG}} \over
V}
\end{equation}
where $L$ is the thickness of the defect and
\begin{equation}
{{n_b^{(0)}} \over s} = {1 \over {4 \pi^2}} \kappa \alpha_W^4 (g^*)^{-1}
\Delta \theta_{CP} ({{m_l} \over T})^2 {{m_H} \over {\lambda_D}} {{\xi^L} \over
{D_L}}.
\end{equation}
The diffusion constant is proportional to $\alpha_W^{-2}$ (see Ref. \cite{JPT2
94}):
\begin{equation}
{1 \over {D_L}} \simeq 8 \alpha_W^2 T.
\end{equation}
Hence, provided that sphalerons do no equilibrate in the diffusion tail,
\begin{equation}
{{n_b^{(0)}} \over s} \sim 0.2 \alpha_W^2 (g^*)^{-1} \kappa
\Delta \theta_{CP} {1 \over {v_D}} ({{m_l} \over T})^2 {{m_H} \over T} {{\xi^L}
\over {D_L}}.
\end{equation}
Since $\xi^L / D_L$ is of the order $1 / (T \delta)$, the baryon to entropy
ratio obtained by nonlocal baryogenesis is proportional to $\alpha_W^2$  and
not $\alpha_W^4$ as the result for local baryogenesis.

Now let us compare the effects of top quarks scattering off the interior of the
advancing wall of the defect \cite{CKN2 92}.  Several effects tend to decrease
the contribution of the top quarks relative to that of tau leptons. Firstly,
their contribution is suppressed \cite{GS 94} since the diffusion tail is cut
off in front of the wall by strong sphaleron effects. Secondly, the diffusion
length for top quarks is smaller, thus reducing the volume in which
baryogenesis takes place.  However, there are also enhancement factors, e.g.
the ratio of the squares of the masses $m_t^2/m_{\tau}^2$ (see Eq. 2).

Let us now turn briefly to the case when the walls of the defect are thick in
the sense that $\delta>l$. Then there is no significant reflection of particles
from the defect walls because the particles see an adiabatically changing Higgs
field. One might therefore think that there is no nonlocal baryogenesis in this
case.

However, as has been shown in Ref. \cite{JPT3 94} (see also Ref. \cite{CKN
94}), in the case of thick walls there is a classical force in the equilibrium
equations for baryon number that drives the equilibrium value away from zero,
generating a net baryon asymmetry.
We shall not consider this mechanism in detail here but shall just note that
whatever the configuration or relative dimensions of the defects we always
produce some contribution to the baryon asymmetry from nonlocal processes.

\section{Specific Geometries and Examples}
Now that we have examined in detail the various ways in which an excess of
baryons may be generated by a topological defect we can compute the total
baryon asymmetry of the universe for a given type of defect with a given
distribution.

Let us assume that in a volume $V=x^3$ there is a single defect of a given type
with the electroweak symmetry restored out to a distance $R_s$. The scale $x$
may be considered as the mean separation of defects. Then, for a single defect
the volume suppression factor is

\begin{eqnarray}
\frac{V_{\rm BG}}{V} & \simeq & \pi \frac{R^2R_s}{x^3} \ \ \ \ \ \mbox{String
loop, Radius R} \\
                     & \simeq & \frac{R_s}{x}v_D       \ \ \ \ \
\mbox{``Infinite" String} \\
                     & \simeq & {\cal O}(1)v_D        \ \ \ \ \
\mbox{``Infinite" Domain Wall} \\
                     & \simeq & \frac{4\pi}{3}\left(\frac{R}{x}\right)^3 \ \ \
\ \mbox{Domain Bubble, Radius R} \\
                     & \simeq & 4\pi\left(\frac{R}{x}\right)^2 v_D \ \ \ \
\mbox{Translating stable bubble}
\end{eqnarray}
Since we are including the total volume swept by the defects we also need to
consider the possible additional suppression from the factor ${\bar \Gamma}_s
L/v_{\rm D}$ due to all the antibaryons not having time to decay before the
trailing edge of the defect passes.

\subsection{Local and Nonlocal Baryogenesis from Cosmic Strings}
In our original analysis \cite{BDT 94} we concentrated on the contribution to
the baryon asymmetry from local baryogenesis in loops of cosmic string produced
by a symmetry breaking at a scale $\eta>\eta_{\rm EW}$. Thus, it is interesting
to compare the results of that calculation with our new predictions.

We shall consider two possibilities

\begin{enumerate}
\item The scale $\eta$ is sufficiently close to the electroweak scale that the
string network is in the friction dominated epoch at the time of the
electroweak phase transition.
\item $\eta$ is sufficiently greater than $\eta_{\rm EW}$ that the string
network has reached a scaling solution by $t_{\rm EW}$. Note that strings with
a mass per unit length $\mu$ remain in the friction dominated era until
\cite{KEH} a time
$t^* = (G \mu)^{-1} t_c$, where $t_c$ is the time corresponding to the critical
temperature of the phase transition.
\end{enumerate}
In the first case we will make the approximation that all string loops have the
same radius but in the second case we shall integrate over the loop
distribution function.

Firstly assume that the network is in the friction dominated epoch at $t_{\rm
EW}$. Note that in this case (except in a narrow window for $\eta$)
superconducting strings do not satisfy equation (15) and the strings are
therefore (for local baryogenesis) thin enough that the additional suppression
factor ${\bar \Gamma}_s L/v_{\rm D}$ mentioned above applies in both the
ordinary and superconducting cases (for a large range of Higgs self-coupling).
For nonlocal baryogenesis the suppression factor is linear in $L v_D / D$. Let
us further assume that we have one string loop per correlation volume at
formation, via the Kibble mechanism. In one horizon volume per expansion time
the total volume taking part in baryogenesis is
\begin{equation}
V_{\rm BG} = R_s\xi(t)^2\left(\frac{t}{\xi(t)}\right)^3v_D
\end{equation}
where we have used the largest strings with radius equal to the correlation
length $\xi(t)$ and the last factor is the number of string loops per horizon
volume. Thus, dividing by the horizon volume $t^3$ we obtain the volume
suppression factor
\begin{equation}
(SF) = \frac{V_{\rm BG}}{V} = \frac{R_s}{\xi(t)}
\end{equation}
Using $\xi(t_f)\simeq \lambda^{-1}\eta^{-1}$ \cite{K 76} where $t_f$ is the
formation time of the string network and \cite{KEH}
\begin{equation}
\xi(t) \sim \xi(t_f)\left(\frac{t}{t_f}\right)^{5/4}
\end{equation}
we obtain (SF) as
\begin{eqnarray}
(SF) & = & \lambda\left(\frac{\eta_{\rm EW}}{\eta}\right)^{3/2}v_D\ \ \ \
\mbox{Ordinary Strings} \\
     & = & \lambda\left(\frac{\eta_{\rm EW}}{\eta}\right)^{3/4}v_D\ \ \ \
\mbox{Superconducting Strings}
\end{eqnarray}
These equations take into account only the dynamics during the first Hubble
expansion time after $t_{EW}$. In later expansion times, the density of strings
is diluted, and hence the above results are a good approximation of the total
effect of strings.

The above suppression factors are an improvement over the original ones by a
factor of 1/2 in the exponent and mean that the phase transition giving rise to
the neccessary defects need not lie quite so close to the electroweak scale as
once imagined.

Let us now turn to the case where the defects are formed at a scale much higher
than the electroweak scale so that the defect network is well described by a
scaling solution at $t_{\rm EW}$. If the strings are ordinary we expect still
to have the ${\bar \Gamma}_s L/v_{\rm D}$ suppression but for superconducting
defects we shall see that this is absent since the electroweak symmetry is
restored out to such a large radius that all the antibaryons may decay before
the baryons are created.

Let us again focus on string loops. The number density of string loops with
radii in the range $[R,R+dR]$ is given by \cite{ZV}
\begin{equation}
n(R,t) = \left\{ \begin{array}{ll}
         \nu R^{-5/2} t^{-3/2}    & \ \ \ \mbox{$\gamma t < R < t$} \\
         \nu \gamma^{-5/2} t^{-4} & \ \ \ \mbox{$R < \gamma t$}
         \end{array} \right.
\end{equation}
where $\gamma \ll 1 $ is a constant determined by the strength of
electromagnetic radiation from the string.  Loops with radius $R=\gamma t$
decay in one Hubble expansion time.  In the above we are assuming that
electromagnetic radiation dominates over gravitational radiation.  If this is
not the case, then $\gamma$ must be replaced by $\gamma_g \, G \mu $, $\mu$
being the mass per unit length of the string $(\mu \simeq\eta^2)$ and \cite{VV}
$\gamma_g \sim 100$. In other words, $\gamma$ is bounded from below
\begin{equation}
\gamma > \gamma_g G \mu\, .
\end{equation}

We can estimate the suppression factor (SF) by integrating over all the string
loops present at $t_{\rm EW}$
\begin{equation}
(SF)\simeq \pi \int_0^{\gamma t_{EW}} \, d R \, R^2\, R_s \, n \left( R,t_{EW}
\right) = \frac{\pi}{3} \, \nu\, \gamma^{1/2} \left(\frac{R_s}{t_{EW}}\right)
\end{equation}
Without superconductivity the suppression factor for GUT strings ($\eta =
10^{16}$ GeV) is so small ($\sim 10^{-32}$) that the contribution is
negligible. However, with superconducting strings we may estimate
\begin{equation}
(SF) \sim \nu \gamma^{1/2}\left(\frac{\eta}{m_{pl}}\right)
\end{equation}
so that the final baryon to entropy ratio generated by this mechanism is given
by
\begin{equation}
\frac{n_B}{s}=\frac{n^0_B}{s}(SF)
\end{equation}
with (SF) given by the above and $n^0_B/s$ proportional to $\alpha_W^4$ for
local baryogenesis and to $\alpha_W^2$ for nonlocal baryogenesis.
Clearly, with the volume enhancement and nonlocal effects, this is an
improvement over our original mechanism but unfortunately still lies below the
observed value.

\subsection{Nonlocal Barogenesis from Cosmic Domain Walls}
Let us now briefly examine what is probably the best case scenario for our
mechanism.

Assume that in the early universe there is a symmetry breaking at a scale
$\eta$ such that cosmic domain walls are formed. One important caveat, however,
is that we must assume the existence of some process by which the domain walls
are removed at a later time so that they do not come to dominate the energy
density of the universe.

Let us first focus on domain walls formed at a scale close to the electroweak
scale so that they are in the friction dominated epoch at the time of the
electroweak phase transition. We consider the effect of ``infinite" walls in
which case $(SF) \sim v_D$. Clearly, for a wide range of domain wall velocities
$v_D$ the resulting baryon to entropy ratio is comparable with what results
from first order mechanisms and can agree with the observed baryon to entropy
ratio for suitable choices of the parameters.

{}From equation~(25) we may estimate the relevant quantities from Ref.
\cite{JPT3 94} and arrive at
\begin{equation}
{{n_b^{(0)}} \over s} \sim 10^{-6}\kappa\Delta\theta_{\rm CP}y_{\tau}^2v_D
\end{equation}
where $y_{\tau}$ is the Yukawa coupling for $\tau$-leptons.

Note that if the scaling solution for domain walls is maintained long after
$t_{EW}$, then the contributions from different Hubble time steps add up and
can give an additional enhancement of the topological defect-mediated
baryogenesis scenario.

Imagine that the scaling solution for domain walls lasts for sufficiently many
expansion times that an equilibrium baryon number is reached. This means that
an equal number of baryons will be created as destroyed in the passage of a
wall. If the equilibrium value for baryon number is denoted as $B_0$, the
equilibrium balance equation, obtained by considering the trailing edge of the
wall, becomes
(for local baryogenesis)
\begin{equation}
B_0 e^{-{{{\bar \Gamma_s} L} \over {v_D}}} + n_b^0 (1 - e^{-{{{\bar \Gamma_s}
L} \over {v_D}}}) = B_0
\end{equation}
The first term is the baryon density left over from what enters the leading
edge of the wall, the second term is what is created in front of the trailing
edge. For nonlocal baryogenesis the balance equation reads
\begin{equation}
B_0 e^{-{{{\bar \Gamma_s} L} \over {v_D}}} + n_b^0 (1 - e^{-{{{L v_D} \over
D}}}) = B_0,
\end{equation}
which gives
\begin{equation}
B_0 = n_b^0
\end{equation}
for local baryogenesis, and
\begin{equation}
B_0 = n_b^0 {{v_D^2} \over {{\bar \Gamma_s} D}}
\end{equation}
for nonlocal baryogenesis (where we have assumed that the core is thin). Note
that this result is unsuppressed for local baryogenesis and in fact enhanced
for the nonlocal mechanisms (see e.g. \cite{JPT2 94}). The interesting fact is
that in the nonlocal case, the dependence on the sphaleron rate drops out of
the final result completely. In order to obtain equilibrium, the number of
expansion times during which the scaling solution persists must be larger than
${1 \over {{\bar \Gamma_s} L}}$.

\section{Discussion and Conclusions}
We have investigated in detail the production of a baryonic asymmetry at the
electroweak scale where the departure from thermal equilibrium is realized by
the motion of topological defects remaining after a previous phase transition.
The main advantage of this scenario is that it is insensitive to the details,
including the order, of the electroweak phase transition and is still effective
even if sphalerons are in thermal equilibrium just below the phase transition
temperature.

The electroweak symmetry is restored out to some distance around these defects
and, as points in space make the transition from true to false vacuum and back
again as the defect passes by, CP violation in the walls of the defect results
in the production of a net baryonic excess.

Our analysis addresses the qualitatively different mechanisms of local and
nonlocal baryogenesis and in each case we have evaluated the possible baryon to
entropy ratio which may be generated by defects of a given dimension and
distribution. In particular we have addressed the case where the defects are
cosmic string loops. The key observation in this paper is that the effective
volume contributing to baryogenesis is much more than was assumed at first in
Ref.~1. For the scenario with cosmic string loops this was shown to result in a
less severe suppression than originally calculated.

Further, we have included in our calculation of the nonlocal mechanism the
effects of particle transport. This allows us to make a detailed estimate of
the total baryon asymmetry due to the imperfect cancellation of the baryons and
antibaryons produced on opposite faces of the defect.

It may be useful to summarize the conditions under which our approximations in
the analysis of local and nonlocal baryogenesis apply.

First, a sphaleron has to fit within the defect core (nonlocal BG) or wall
(local BG). From the sphaleron rate $\Gamma_s$ in the unbroken phase (see (6)),
we can infer that the sphaleron radius is smaller than $(\alpha_W T)^{-1}$. In
the walls, the sphalerons are rather small ($ \sim m_W^{-1}$). Hence, this
condition is
\begin{equation}
L < (\alpha_W T)^{-1} \ , \ \ \ \delta < m_W^{-1},
\end{equation}
where $L$ and $\delta$ are defect core and wall radii respectively.

As discussed in Section 3, in the case of local baryogenesis, antibaryons
produced at the leading edge of the defect can annihilate with the baryons
produced at the trailing edge unless the baryon density equilibrates to zero in
the defect core via sphaleron processes. For this to occur, the defect must be
sufficiently thick:
\begin{equation}
{L \over {v_D}} > {\bar \Gamma}_s^{-1},
\end{equation}
where $v_D$ is the transverse velocity of the defect. If this condition is not
satisfied, there will be a suppression of the baryon to entropy ratio linear in
${\bar \Gamma}_s L / v_D$.

Lastly, in the case of nonlocal baryogenesis, baryon production takes place in
the diffusion tail which extends in front of the trailing defect edge. In order
not to get a suppression of the effect, the core must be thick compared to the
diffusion tail $D / v_D$:
\begin{equation}
{{L v_D} \over D} > 1.
\end{equation}

In calculating the effects of nonlocal baryogenesis we have used the reflection
coefficients of leptons from defects calculated for planar walls. The
reflection is dominated by particles with wavelength $\lambda \sim \delta$.
Hence, a condition for the applicability of our calculations is $\lambda <<
R_s$, where $R_s$ is the curvature radius of the defect wall. For domain walls,
$R_s >> L$, whereas for strings $R_s \sim L$.

In nonlocal baryogenesis mediated by a classical force \cite{JPT3 94} there
are, as in the case of local baryogenesis, no further geometric suppression
factors.
However, for nonlocal baryogenesis by quantum reflection \cite{{CKN2 92},{JPT1
94}} there is a further condition. When averaged over phase space, the incident
angle of the fermions which scatter off the defect wall is peaked at a value of
\begin{equation}
{{m_H} \over {2T}} \simeq {1 \over {T \delta}},
\end{equation}
where $m_H$ is the Higgs mass which determines the wall thickness. In order for
the single scattering calculations used in this paper to be valid, the typical
distance $d$ the fermions travel within the core after a reflcetion before
hitting the wall a second time, which is
\begin{equation}
d \simeq {{2 R_s} \over {T \delta}},
\end{equation}
must be larger than the diffusion length $6 D$, i.e.
\begin{equation}
d > 6 D.
\end{equation}
This condition is satisfied provided that the wall thickness is significantly
smaller than the defect core size. If it is violated, then there will be a
further suppression factor.
\vspace{1cm}
\begin{center}
\bf Acknowledgements
\end{center}
\vspace{5mm}
We would like to thank Neil Turok and Michael Joyce for useful discussions.
This work was supported in part by the US Department of Energy under Grant
DE-FG0291ER40688, Task A and by an NSF-SERC Collaborative Research Award
NSF-INT-9022895 and SERC GR/G37149. T.P. is partially supported by NSF contract
PHY90-21984 and the David and Lucile Packard Foundation.

\end{document}